\documentclass[journal]{IEEEtran}
\usepackage{subcaption}
\usepackage{framed,multirow}
\usepackage{cite}
\usepackage{amssymb}
\usepackage{latexsym}

\usepackage{url}
\usepackage{xcolor}

\usepackage{hyperref}

\usepackage{mathtools,marvosym}
\usepackage{color}
\usepackage{graphicx} 
\usepackage{multicol}
\usepackage{tcolorbox}
\usepackage[]{xcolor}
\usepackage{colortbl}

\renewcommand{\tablename}{Table}

\definecolor{pupil}{RGB}{0,137,255}
\definecolor{tape}{RGB}{255,165,0}
\definecolor{retractors}{RGB}{99,0,255}
\definecolor{iris}{RGB}{255	,0,	0}
\definecolor{skin}{RGB}{255	,0,	165}
\definecolor{cornea}{RGB}{255	,255,	255}
\definecolor{hcannula}{RGB}{141	,141,	141}
\definecolor{vcannula}{RGB}{255	,218,	0}
\definecolor{capcyst}{RGB}{173	,156,	255}
\definecolor{rcannula}{RGB}{73	,73,	73}
\definecolor{bonn}{RGB}{250	,213,	255}
\definecolor{primknife}{RGB}{255	,156,	156}
\definecolor{phaco}{RGB}{99	,255,	0}
\definecolor{lensinjector}{RGB}{157,	225	,255}
\definecolor{iahandpiece}{RGB}{255,	89	,124}
\definecolor{secondknife}{RGB}{173,	255	,156}
\definecolor{micromanipulator}{RGB}{255,	60	,0}
\definecolor{iahandle}{RGB}{40	,0,	255}
\definecolor{capforceps}{RGB}{170,	124	,0}
\definecolor{capcysthandle}{RGB}{0,	255,207}
\definecolor{hand}{RGB}{255,156,201}
\definecolor{secondknifehandle}{RGB}{188,0,255}
\renewcommand{\tablename}{Table}
\DeclareRobustCommand*{\IEEEauthorrefmark}[1]{%
    \raisebox{0pt}[0pt][0pt]{\textsuperscript{\footnotesize\ensuremath{#1}}}}

\begin{document}
\title{2020 CATARACTS Semantic Segmentation Challenge}

\author{%\IEEEauthorblockN{Author One\IEEEauthorrefmark{1},
%Author Two\IEEEauthorrefmark{2}, Author Three\IEEEauthorrefmark{2} and
%Author Four\IEEEauthorrefmark{1}\\}
\IEEEauthorblockN{Imanol Luengo\IEEEauthorrefmark{1}, Maria Grammatikopoulou\IEEEauthorrefmark{1}, Rahim Mohammadi\IEEEauthorrefmark{1}, Chris Walsh\IEEEauthorrefmark{1}, Chinedu Innocent Nwoye\IEEEauthorrefmark{2}, Deepak Alapatt\IEEEauthorrefmark{2}, Nicolas Padoy\IEEEauthorrefmark{2}, Zhen-Liang Ni\IEEEauthorrefmark{3}, Chen-Chen Fan\IEEEauthorrefmark{3}, Gui-Bin Bian\IEEEauthorrefmark{3}, Zeng-Guang Hou\IEEEauthorrefmark{3}, Heonjin Ha\IEEEauthorrefmark{4}, Jiacheng Wang\IEEEauthorrefmark{5}, Haojie Wang\IEEEauthorrefmark{5}, Dong Guo\IEEEauthorrefmark{6}, Lu Wang\IEEEauthorrefmark{6}, Guotai Wang\IEEEauthorrefmark{6}, Mobarakol Islam\IEEEauthorrefmark{7}, Bharat Giddwani\IEEEauthorrefmark{8}, Ren Hongliang\IEEEauthorrefmark{7}, Theodoros Pissas\IEEEauthorrefmark{9,10}, Claudio Ravasio\IEEEauthorrefmark{9,10}, Martin Huber\IEEEauthorrefmark{9}, Jeremy Birch\IEEEauthorrefmark{9}, Joan M. Nunez Do Rio\IEEEauthorrefmark{11}, Lyndon da Cruz\IEEEauthorrefmark{12}, Christos Bergeles\IEEEauthorrefmark{9,11,12}, Hongyu Chen\IEEEauthorrefmark{13}, Fucang Jia\IEEEauthorrefmark{13}, Nikhil Kumar Tomar\IEEEauthorrefmark{14}, Debesh Jha\IEEEauthorrefmark{14}, Michael A. Riegler\IEEEauthorrefmark{14}, Pal Halvorsen\IEEEauthorrefmark{14}, Sophia Bano\IEEEauthorrefmark{10,15}, Uddhav Vaghela\IEEEauthorrefmark{10,15}, Jianyuan Hong\IEEEauthorrefmark{16}, Haili Ye\IEEEauthorrefmark{16}, Feihong Huang\IEEEauthorrefmark{16}, Da-Han Wang\IEEEauthorrefmark{16}, and Danail Stoyanov\IEEEauthorrefmark{1,10} \\}
\vspace{0.5cm}
\IEEEauthorblockA{
\IEEEauthorrefmark{1}Digital Surgery Ltd, Medtronic 
\IEEEauthorrefmark{2}ICube, University of Strasbourg, CNRS, IHU Strasbourg, France
\IEEEauthorrefmark{3}Institute of Automation, Chinese Academy of Sciences
\IEEEauthorrefmark{4}Hutom, Seoul, South Korea
\IEEEauthorrefmark{5}Department of Computer Science, Xiamen University, Xiamen, China
\IEEEauthorrefmark{6}University of Electronic Science and Technology of China, Chengdu, China
\IEEEauthorrefmark{7}Dept. of Biomedical Engineering, National University of Singapore
\IEEEauthorrefmark{8}Dept. of Electronics and Telecoms. Engineering, National Institute of Technology, Raipur, India
\IEEEauthorrefmark{9}School of Biomedical Engineering \& Imaging Sciences, King’s College London, London, UK
\IEEEauthorrefmark{10}Wellcome/EPSRC Centre for Interventional and Surgical Sciences (WEISS), University College London, London, UK
\IEEEauthorrefmark{11}Institute of Ophthalmology, University College London, London, UK
\IEEEauthorrefmark{12}Moorfields Eye Hospital, London, UK
\IEEEauthorrefmark{13}Shenzhen  Institute  of  Advanced  Technology,  Chinese  Academyof  Sciences
\IEEEauthorrefmark{14}SimulaMet, UiT The Arctic University of Norway, Oslo Metropolitan University
\IEEEauthorrefmark{15}Department of Computer Science, University College London, UK
\IEEEauthorrefmark{16}School  of  Computer  and  Information Engineering, Xiamen University of Technology, China
}
}
\maketitle

\begin{abstract}
Surgical scene segmentation is essential for anatomy and instrument localization which can be further used to assess tissue-instrument interactions during a surgical procedure. In 2017, the Challenge on Automatic Tool Annotation for cataRACT Surgery (CATARACTS) released 50 cataract surgery videos accompanied by instrument usage annotations. These annotations included frame-level instrument presence information. In 2020, we released pixel-wise semantic annotations for anatomy and instruments for 4670 images sampled from 25 videos of the CATARACTS training set. The 2020 CATARACTS Semantic Segmentation Challenge, which was a sub-challenge of the 2020 MICCAI Endoscopic Vision (EndoVis) Challenge, presented three sub-tasks to assess participating solutions on anatomical structure and instrument segmentation. Their performance was assessed on a hidden test set of 531 images from 10 videos of the CATARACTS test set.
\end{abstract}

\section{Introduction}
Surgical video processing could facilitate pre-operative planning, intra-operative image guidance and generation of post-operative analysis of the surgical procedure. Computer-assisted interventions have the potential to enhance the surgeon's visualization and navigation capabilities and post-operative analytics to provide insights for surgical training and risk assessment. A necessary element for these processes is scene understanding and, in particular, anatomy and instrument detection and localization. Therefore, by segmenting and differentiating among the elements that appear in the surgical view, it is possible to assess tissue-instrument interactions and understand surgical workflow.
\begin{figure}[t]
\centering
\includegraphics[width=\linewidth]{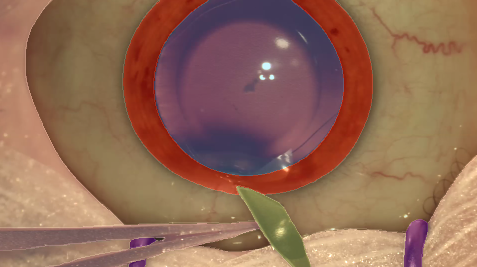}
\caption{Example frame with overlaid reference mask from the dataset used in the challenge. (Colormap: \textcolor{pupil}{\rule{0.5cm}{0.25cm}} Pupil, \textcolor{iris}{\rule{0.5cm}{0.25cm}} Iris,  $^{\framebox{ \textcolor{cornea}{\rule{0.3cm}{0.001cm}}}}$ Cornea, \textcolor{skin}{\rule{0.5cm}{0.25cm}} Skin, \textcolor{tape}{\rule{0.5cm}{0.25cm}} Surgical tape, \textcolor{retractors}{\rule{0.5cm}{0.25cm}} Eye retractors,
\textcolor{bonn}{\rule{0.5cm}{0.25cm}} Bonn Forceps,
\textcolor{secondknife}{\rule{0.5cm}{0.25cm}} Secondary Knife)}
\label{tab:seg_reference}
\end{figure}
\begin{figure*}[t]
\centering
\includegraphics[width=\linewidth]{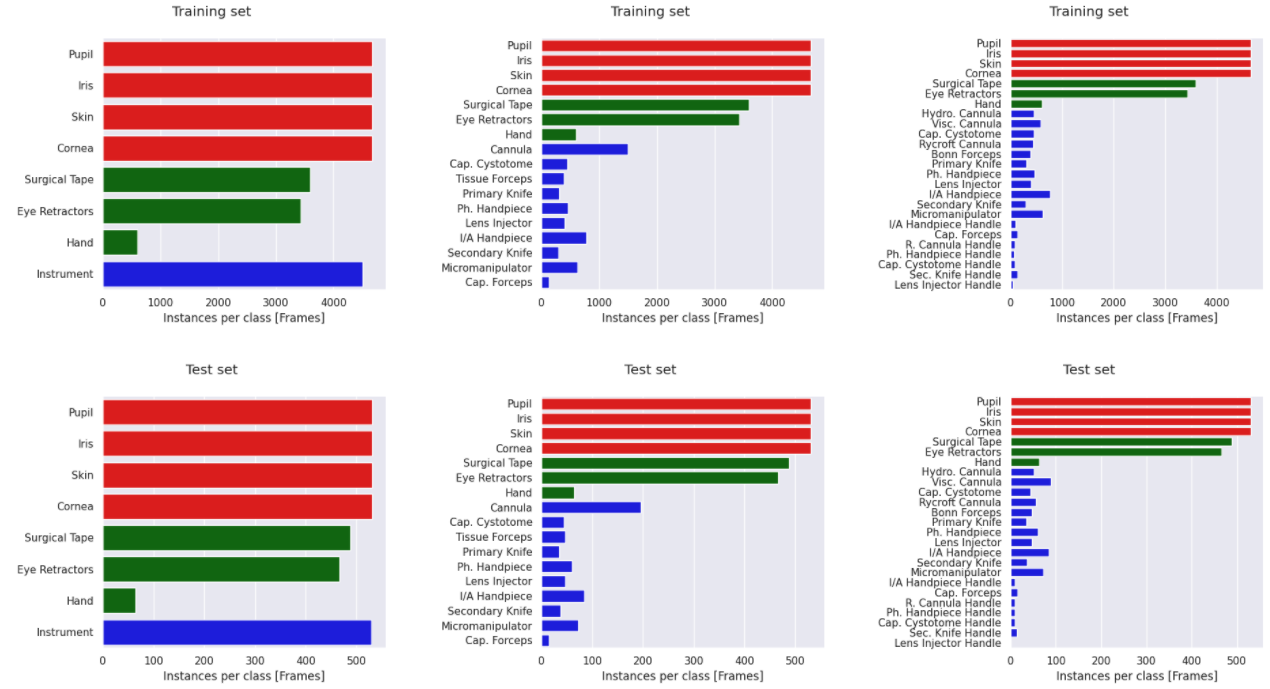}
\caption{Classes and their distributions for the training and test set for Task I (first column), Task II (second column) and Task III (third column)}
\label{fig:task-dist}
\end{figure*}
Recent advances in the fields of computer vision and deep learning have accelerated the development of segmentation and detection models. However, these algorithms require large amounts of labelled data for training and to assess their efficiency. In particular, pixel-wise annotation is necessary for the development of image segmentation algorithms, a laborious and time-consuming task that is a significant overhead of these algorithms. Therefore, the research community is lacking of sufficient amount data, and in particular labelled data, to provide solutions. 

Previously, the surgical computer vision community has organized various challenges towards providing and assessing solutions to the above mentioned clinical and technical problems. In 2017, the Challenge on Automatic Tool Annotation for cataRACT Surgery (CATARACTS) \footnote{https://cataracts.grand-challenge.org/} was organized and its task was instrument detection in microscope videos from cataract surgeries. The dataset included 50 videos (25 for training and 25 for testing) with frame-level instrument annotations \cite{cataracts}. Additionally, in 2017 and 2018, the Endoscopic Vision (EndoVis) challenge included sub-challenges on instrument detection and segmentation in robotic surgery \cite{endovis2017,allan20202018}. One important contribution of these challenges is the public release of labelled data that is vital to develop deep learning algorithms by the research community. 

The previous EndoVis sub-challenges focused on the detection and segmentation of robotic instruments in endoscopic videos. In 2018, the dataset of \cite{endovis2017} was extended to include some anatomical structures that are apparent in the scene. However, assessing anatomy and instrument segmentation in other surgical procedures which include a larger variety and less standardized range of instruments present additional problems to tackle. While issues such as accurate and spatially consistent instrument segmentation is a common problem for robotic, laparoscopic or instruments used in other procedures, the instruments used in cataract surgery can come from different manufacturers with different appearances for the same type of instrument and variable sizes among instruments. In addition, it is also important to segment anatomical structures to understand interactions in the scene.

To this end, the 2020 CATARACTS Semantic Segmentation Challenge, which was a sub-challenge of the 2020 MICCAI EndoVis Challenge, presented three sub-tasks to assess participating solutions on anatomical structure and instrument segmentation in cataract surgery videos. The pixel-wise semantic annotations provided for training and testing were generated from the videos released in the 2017 CATARACTS tool detection Grand Challenge.
 
\section{Challenge Overview}
The goal of the challenge was to assess solutions for scene segmentation in cataract surgery videos. One of the problems to tackle in this dataset is class imbalance as instruments appear less frequently in the scene than anatomy and occupy smaller regions in the frame than the anatomical structures. Therefore, three different sub-tasks were defined in order to assess different levels of imbalance and its impact on semantic segmentation. A number of 9 to 11 teams participated across the three tasks. The challenge allowed teams to either use state-of-the-art models or improve on existing solutions. In the following sections, a presentation of the dataset is given, along with the definition of the three sub-tasks and overall challenge organization and setup.
\begin{figure}[t!]
\centering
\includegraphics[width=\linewidth]{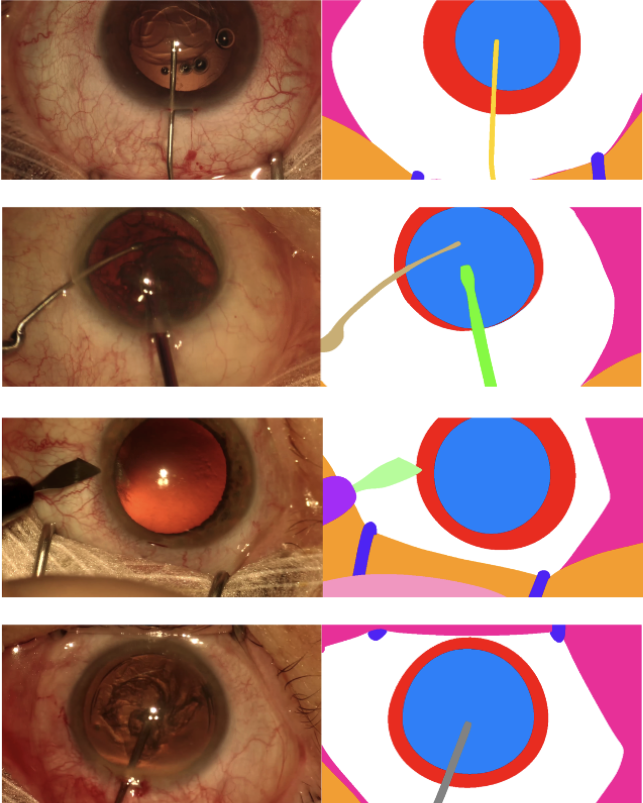}
\caption{Example frames and reference segmentation masks from the training (first and second row) and test set (third and fourth row). (Colormap: \textcolor{pupil}{\rule{0.5cm}{0.25cm}} Pupil, \textcolor{iris}{\rule{0.5cm}{0.25cm}} Iris,  $^{\framebox{ \textcolor{cornea}{\rule{0.3cm}{0.001cm}}}}$ Cornea, \textcolor{skin}{\rule{0.5cm}{0.25cm}} Skin, \textcolor{tape}{\rule{0.5cm}{0.25cm}} Surgical tape, \textcolor{retractors}{\rule{0.5cm}{0.25cm}} Eye retractors,
\textcolor{hand}{\rule{0.5cm}{0.25cm}} Hand,  
\textcolor{vcannula}{\rule{0.5cm}{0.25cm}} Viscoelastic cannula, 
\textcolor{micromanipulator}{\rule{0.5cm}{0.25cm}} Micromanipulator,  \textcolor{phaco}{\rule{0.5cm}{0.25cm}} Phacoemulsification handpiece, 
\textcolor{secondknife}{\rule{0.5cm}{0.25cm}} Secondary knife,  
\textcolor{secondknifehandle}{\rule{0.5cm}{0.25cm}} Secondary knife handle,
\textcolor{hcannula}{\rule{0.5cm}{0.25cm}} Hydrodissection cannula)}
\label{fig:examples-train-test}
\end{figure}
\subsection{Dataset}
The challenge dataset includes 4670 training images and 531 test images sampled from the 25 videos of the CATARACTS training and from 10 videos of the test set respectively. The annotation process was as follows: First, a team of clinical experts drafted the guidelines for anatomy and instrument annotation. The guidelines were later handed to an internal team of annotators who were briefed on the guidelines and the cataract surgery procedures. Each frame was annotated by one person and checked by a second one. Cases of disagreement in annotation were resolved by the clinical team. Further pixel-wise checks were performed to ensure accurate borders and correct pixel-wise labelling.
\subsubsection{Training data}
The training set consists of 4670 images with semantic labels of 36 classes and, in particular, 4 anatomical classes, 29 instruments and 3 classes of other objects appearing in the scene. The images are sampled from 25 videos of the CATARACTS Challenge training set. More detailed analysis about the training dataset can be found in \cite{grammatikopoulou2019cadis}.
\subsubsection{Test set}
The test set includes 531 labelled images sampled from 10 videos of the CATARACTS challenge test set. In total, 25 classes are present in the test set and in particular, 4 anatomical classes, 18 instruments and 3 other objects in the scene. It is worth noting that the classes from the training set that were ignored in the three sub-tasks were not selected to be present in the test set. The test set was not released to the participants.
\subsection{Challenge tasks}
The three sub-tasks aim at representing different use cases of scene segmentation and also highlight the technical issues appearing in segmentation and instrument classification. 
\subsubsection{Task I - Anatomy and instrument}
The goal for the first task was to focus on segmentation through mitigating the imbalance issue. To this end, all instrument classes were merged into a single class while all anatomy and other objects remain as separate classes. Therefore, 8 classes were identified for this task: 4 for anatomy (Pupil, Iris, Cornea and Skin), 3 for other objects (Surgical Tape, Eye retractors and Hand) and 1 for all instruments. The class distributions for the training and test set are depicted in Figure \ref{fig:task-dist}. It can be seen that by merging instrument classes the distributions become more balanced for both training and test sets.
\subsubsection{Task II - Anatomy and grouped instruments}
The second task includes 17 classes and, in particular, 4 anatomical classes, 3 of other objects and 10 instruments (Figure \ref{fig:task-dist}). In this task, the instruments were merged according to appearance and usage as suggested by the clinical experts who drafted the annotation guidelines. More specifically, all cannulas were merged and represented by one class, the Bonn and Troutman forceps were merged as Tissue forceps and all instruments and tips were merged as one instrument class. Classes with insufficient instances in the training set were ignored in case they could not be merged with another instrument class. These include: Suture Needle, Needle holder, Vitrectomy Handpiece, Mendez Ring, Marker, Cotton, Iris Hooks. The class distributions for this task are shown in Figure \ref{fig:task-dist}.
\begin{figure}[t!]
\centering
\includegraphics[width=\linewidth]{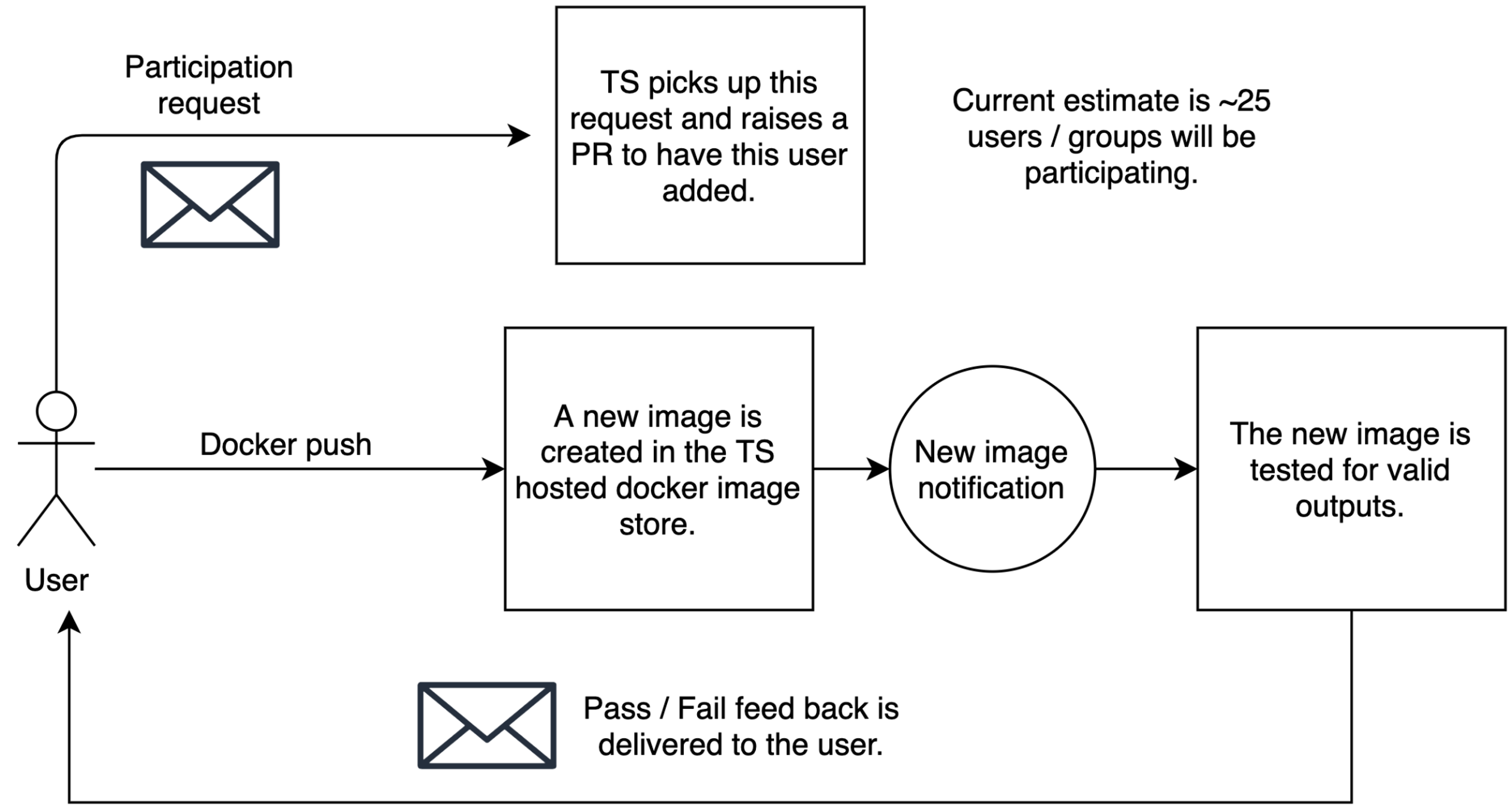}
\caption{Challenge setup and workflow for docker container submission}
\label{fig:challenge-setup}
\end{figure}
\begin{figure*}[t!]
\centering
\includegraphics[width=\linewidth]{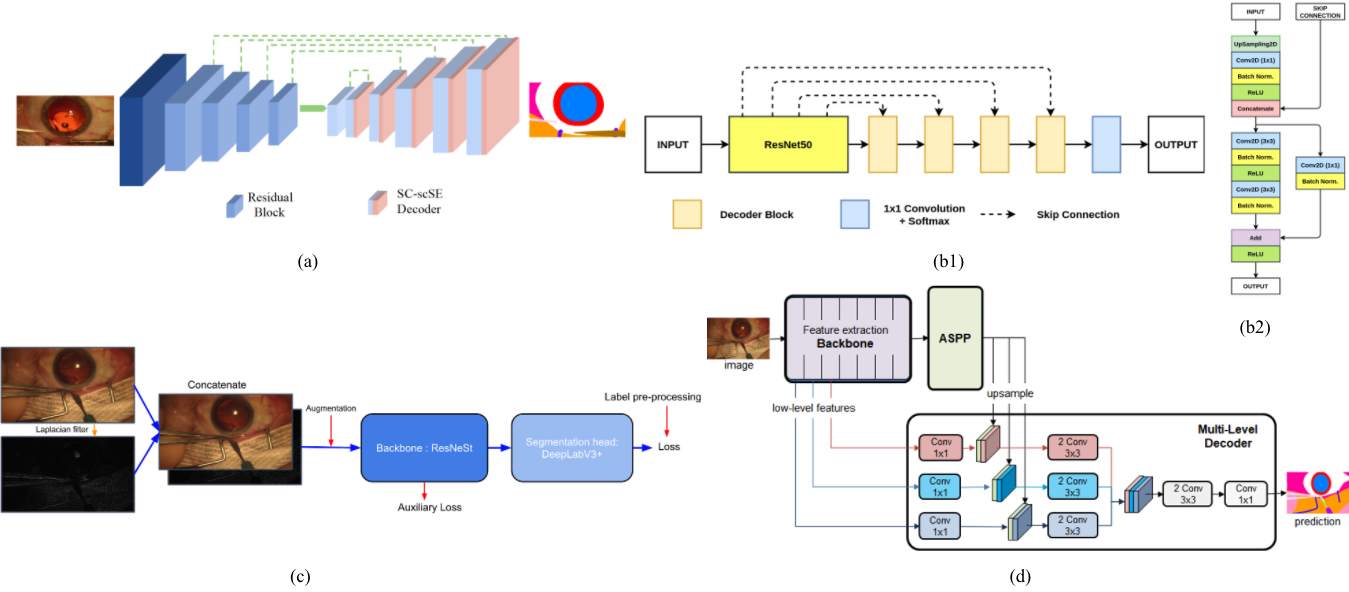}
\caption{Model diagrams for: Perception (a), SimulaMet (b1-2), HUTOM (c) and CAMMA-CADIS (d)}
\label{fig:models-participants-1}
\end{figure*}
\subsubsection{Task III - Anatomy, instrument tips and handles}
The third tasks includes 25 classes (4 anatomical classes, 3 of other objects and 18 instruments). In this tasks, all tips and instrument handles are represented as separate classes while, similarly to task II, instruments with insufficient instances for training were ignored. The ignored classes are: Suture Needle, Needle holder, Charleux Cannula, Vitrectomy Handpiece, Mendez Ring, Marker, Troutman Forceps, Cotton, Iris Hooks, Hydrodissection Cannula Handle and Primary Knife Handle. The class distributions for task III can be seen in Figure \ref{fig:task-dist}.
\begin{comment}
\begin{figure}[t!]
\centering
\includegraphics[width=\linewidth]{challenge_setup.png}
\caption{Challenge setup and workflow for docker container submission}
\label{fig:challenge-setup}
\end{figure}
\begin{figure*}[t!]
\centering
\includegraphics[width=\linewidth]{models-participants.png}
\caption{Model diagrams for: Perception (a), SimulaMet (b1-2), HUTOM (c) and CAMMA-CADIS (d)}
\label{fig:models-participants-1}
\end{figure*}
\end{comment}
\subsection{Challenge organization and setup}
Overall, 48 users expressed interest and registered for the challenge. The timeline of the challenge organization was as follows: A final version of the training set as described in \cite{grammatikopoulou2019cadis} was made publicly available in April 2020. By end of August, a slack channel was created for the registered participants while team registration opened on September 1st, 2020. Model submission was opened on 14th September 2020 when teams received credentials for the docker submission system. The teams could submit multiple images during this time to check their submissions on a set of randomly selected images from the training set. The submission was open for two weeks and after the deadline the last docker image submitted by each team was used for evaluation on the hidden test set. Overall, 11 teams participated in the challenge in which all of them submitted to at least 1 task, while 9 teams submitted to all sub-tasks. A diagram showing the workflow including the submission procedure can be seen in Figure \ref{fig:challenge-setup}.
\begin{comment}
\begin{figure}[t]
\centering
\includegraphics[width=\linewidth]{challenge_setup.png}
\caption{Challenge setup and workflow for docker container submission}
\label{fig:challenge-setup}
\end{figure}
\end{comment}
\section{Participating teams}

\subsection{CAMMA-CADIS}
Team CAMMA-CADIS is Chinedu Innocent Nwoye, Deepak Alapatt and Nicolas Padoy from ICube, University of Strasbourg, France. In their method, they proposed a multi-level decoder network for semantic segmentation that was used for all three tasks. For the encoder, similarly to DeepLabV3+, they used Xception-65 for feature extraction, followed by Atrous Spatial Pyramid Pooling (ASPP) for feature refinement. To recover more spatial details, they built a multiple decoder of levels $L$ = {l$_1$, l$_2$, …l$_N$} that combines features from different stages $S$ = {s$_1$, s$_2$, …,s$_N$} of the feature extractor, where $N$ is the maximum decoding level. They maintained $N$ = 3 for their experiment. For each decoding level, l$_k$, the ASPP output was refined using the low-level features obtained from stage s$_k$ of the encoder as shown in Figure \ref{fig:models-participants-1}-d. In this way, the high level features were decoded at different levels of encoding semantics. To generate the output, they concatenated the multi-level decoded features and applied two 3$\times$3 convolutional layers followed by one 1$\times$1 convolutional layer before generating the class-wise probabilities using the softmax activation. Before training, the inputs were downsampled to a resolution of 270 x 480. Random data augmentations of scale [0.5, 2] and brightness (delta=0.2) are used. The model was trained for 200 epochs using  categorical cross entropy as a loss function and Stochastic Gradient Descent with Momentum as the optimizer. They used a “poly” learning rate policy where the initial learning rate ($7e^{-4}$) was multiplied by \begin{math} \big(1-\frac{\text{iteration}}{\text{max iteration}}\big)^{0.9} \end{math}. They fitted a batch size of 4 on a Quadro P5000 GPU and trained for approximately 36 hours.

\subsection{CASIA SRL}
Team CASIA SRL are Zhen Liang Ni, Chen Chen Fan, Gui Bin Bian and Zeng Guang Hou from the Institute of Automation, Chinese Academy of Sciences. In their method, they improved PAANet \cite{ni2020pyramid} where a dilated ResNet50 was adopted as the encoder. In the last stage of ResNet50 dilated convolution was used with a dilation rate of 2 to expand the receptive field. The last two upsampling layers in the decoder of PAANet were removed. The pyramid upsampling module was improved by deformable convolution to learn deformation features. Specifically, in the pyramid upsampling module, multi-scale features are sent to a 3$\times$3 deformable convolution \cite{zhu2019deformable} after 1$\times$1 standard convolution In this way, the network can capture multi-scale deformation features to adapt to the scale and shape variation of surgical instruments. Besides, the channels of multi scale features fed into the pyramid upsampling module are increased to 48 These improvements can significantly reduce computational costs and maintain accuracy. The focal loss was used to train the network with $\gamma$ equal to 4.

\subsection{HUTOM}
Team HUTOM is Heonjin Ha. They used ResNeSt\cite{zhang2020resnest} as their backbone network and DeepLabV3+\cite{chen2018encoder} as their segmentation head network. They modified the last convolutional layer in the DeepLabV3+ model with reflection padding. When using Laplacian filters for the input, they modified the first convolutional layer in the backbone model with the 4 channel input tensor layer. They ensembled the backbone models with sundry or various number of layers and with or without the Laplacian filtered input. They used different ensemble compositions for each task with each model following as stricture as in Figure \ref{fig:models-participants-1}-c. For the ensemble model, they extracted the model weights from 130$_{th}$ epoch and 140$_{th}$ epoch. The compositions of the ensemble models for each task were as follows: For Task I, they used two models with ResNeSt269 backbones with weights from the 130$_{th}$ and 140$_{th}$ epochs with RGB input only, two models with ResNeSt269 backbones with weights from the 130$_{th}$ and 140$_{th}$ epochs with RGB and Laplacian filtered inputs and two models with ResNeSt200 backbones with weights from the 130$_{th}$ and 140$_{th}$ epochs with RGB input only. For Task II, they used two models with ResNeSt269 backbones with weights from the 130$_{th}$ and 140$_{th}$ epochs and two models with ResNeSt200 backbones with weights from the 130$_{th}$ and 140$_{th}$ epochs, all with RGB input only. Lastly, for Task III, they used two models with ResNeSt269 backbones with weights from the 130$_{th}$ and 140$_{th}$ epochs with RGB input only, two models with ResNeSt200 backbones with weights from the 130$_{th}$ and 140$_{th}$ epochs with RGB and Laplacian filtered inputs and two models with ResNeSt200 backbones with weights from the 130$_{th}$ and 140$_{th}$ epochs with RGB input only. During training, ignoring some labels in an image helped to make the class label distribution more balanced. Moreover, by picking images with low pixel accuracy, they found that there are images with noisy labels such as broken, label-shifted, incorrectly annotated images. They, therefore, ignored such images. However, even without these images, the appearing classes are still imbalanced so predicting the less frequent classes was still within some uncertainty. They further sampled images in the less frequent classes twice in every epoch. They also applied 480x480 random crop from the images of base size 520. After that, they applied augmentations such as cut-out and Gaussian blur. The cut-out augmentation was randomly applied 4 times. The loss function used was pixel-wise cross-Entropy Loss with loss weights. They applied the same loss to the auxiliary output of the backbone network. For optimization, they used Stochastic Gradient Descent (SGD) with polynomial learning rate and 140 epochs. The initial learning rate was 0.02. They experimented with several loss functions in order to address the imbalance issue. Lastly, they used multi-scale inference at 0.8, 1.0, 1.25, 1.75, 2.0 and 2.25 of the original image size for each model.
\subsection{JJJ}
Team JJJ are Jiacheng Wang and Haojie Wang from the Department of Computer Science, Xiamen University, Xiamen, China. Their model used information from frames $t-2$, $t-1$ and $t$ from which they extract features through a ResNet50 encoder. The extracted features from the three frames are aggregated temporally through a convolutional LSTM block. The output of this block is then passed to a Refinet decoder and the output segmentation mask is produced through a dense Atrous Spatial Pyramid Pooling (ASPP) segmentation head. 
\subsection{LUCK}
Team LUCK is Dong Guo, Lu Wang, Shuojue Yang and Guotai Wang from the Healthcare Intelligence Laboratory, University of Electronic and Science Technology of China, Chengdu, China. They participated in Tasks I and Task II with the same method. They based their work on HRNet \cite{wang2020deep} architecture in order to augment the high resolution representation by every paralleled resolution representation. To train their network, they used the ohem cross entropy loss to focus more on less accurate and challenging cases. In addition, they used a 5-fold cross validation including the best and final epoch and test time augmentation including vertical flip and multiple scale, which keeps consistency with training augmentation to improve both generalization and stability of the network.
\subsection{Perception}
Team Perception are Mobarakol Islam, Bharat Giddwani and Ren Hongliang from the Department of Biomedical Engineering, National University of Singapore and the Department of Electronics and Telecommunications Engineering, National Institute of Technology, Raipur, India. They used an encoder-decoder architecture for segmentation. The model was adopted from their previous works \cite{islam2019learning,islam2021st} which contains a residual encoder and a Skip-competitive Spatial and Channel Squeeze \& Excitation (SC-scSE) decoder as shown in Figure \ref{fig:models-participants-1}-a. The encoder is
formed by 5 residual layers as ResNet18 and the corresponding decoding block contains convolution, adaptive batch normalization \cite{li2018adaptive}, SC-scSE, and deconvolution sequentially. The SC-scSE decoder retains weak features, excites strong features and performs dynamic spatial and channel-wise feature recalibration which makes the network capable of better feature learning. They used batch size of 10 for training the proposed model. The model was trained with a learning rate of 0.0001, using the Adam optimizer and the momentum and weight decay set as constant to 0.99 and $10^{-4}$, respectively. The input images were  flipped randomly as a part of augmentation. They followed the same data split as \cite{grammatikopoulou2019cadis} for train and validation.
\subsection{RVIM Lab}
The team from RVIM Lab, Kings College London are Theodoros Pissas, Claudio Ravasio, Martin Huber, Jeremy Birch, Joan M. Nunez Do Rio, Lyndon da Cruz and Christos Bergeles. Their method for all experiments consisted of a bagging ensemble of three separately trained semantic segmentation networks with a classic encoder-decoder structure, the chosen decoders being Deeplabv3+ \cite{chen2018encoder}, OCRNet \cite{yuan2019object}, and UPerNet \cite{xiao2018unified}. The selected encoder backbone for Deeplabv3+ and OCRNet was a dilated ResNet50 with output stride of 8 and a standard ResNet34 for UPerNet. The final segmentation prediction was the mean aggregation of their respective outputs. All networks were trained using the Lovasz-Softmax loss \cite{berman2018lovasz}, a surrogate for the otherwise non-differentiable mIoU, the main metric used for the evaluation on the validation dataset. They found this to produce a significant performance improvement over cross-entropy based loss functions, which can be attributed to the more direct optimisation on an approximation of the mIOU. Empirically, this was validated by the observation that the validation Lovasz-Softmax loss correlated closely with the validation mIoU: the training epoch for which the validation loss reached its minimum was consistently close to the epoch with the maximum mIoU, a desirable property that did not hold for cross-entropy-based losses. For OCRNet, they used a weight of 1 for the loss on the final prediction and a weight of 0.4 for the loss on its intermediate low resolution prediction.  Early experiments showed a wide range of models performing very similarly, despite testing various architectures and training protocols. Following Andrej Karpathy's advice \cite{karpathy}, they performed a thorough data inspection. This revealed 179 frames that were at least partially mislabelled, with tool classes being the most frequently affected. They removed these frames from consideration in both training and validation by adding a Boolean flag in their dataset loader, reducing the overall dataset from 4670 to 4491 records. The training / validation split proposed in \cite{grammatikopoulou2019cadis} was applied to all training runs, resulting in 3490 training records and 1001 validation records after correcting for mislabelled frames. For their final submission, all networks were trained on the full dataset containing 4491 records.  All networks were trained for a maximum of 50 epochs using a batch size of 8, an exponentially decaying learning rate with a starting value of $10^{-4}$ and online data augmentation consisting of random flips and affine transformations. They found class-imbalance in the dataset to be a major challenge, with performance on some classes consistently superior to that of other classes. Addressing this was especially crucial for Tasks II and III which became their main focus. They were able to significantly improve performance on the rarest classes for Deeplabv3+ and OCRNet by using repeat factor sampling \cite{gupta2019lvis} with a frequency threshold of 0.15 and for UperNet by applying an adaptive batch sampling algorithm favouring records containing the highest percentage of the currently worst-performing classes, starting after 15 epochs. For each network in the final ensemble, they independently ablated different training and architectural choices by evaluating on the validation set. The best variant of each network was then chosen to form the ensemble. Amongst the factors evaluated were batch size, learning rate decay type, learning rate restarts, augmentation strength, a number of different oversampling strategies to overcome class imbalance, as well as depth of the encoder/decoder networks themselves.
\subsection{Siatcami}
Team Siatcami are Hongyu Chen and Fucang Jia from Shenzhen Institute of Advanced Technology, Chinese Academy of Sciences. Their method was based on cross-consistency principle \cite{ouali2020semi} and DeepLabV3+ network \cite{chen2018encoder}. The main contribution of this solution is the use of semi-supervised DeepLabV3+ semantic segmentation method to build the model. This method was based on the principle of domain adaptation and used a cross-consistency method to set up perturbation functions in multiple auxiliary decoders to improve the performance of the network backbone composed of the main encoder and the main decoder. From the videos provided for training, 22 videos were selected as the training set, and the other three were used as the validation set. In addition, 3,500 images without ground truth labels randomly chosen from the CATARACTS 2018 were used as the unlabeled data part of the training set. The images were resized to 512×512 pixels as network input, and random scaling and random horizontal flipping were applied as data augmentation. The loss function combined a cross-entropy loss function and a square-difference loss function. The SGD optimizer was used in training with a batch size of 8 for 80 iterations.
\subsection{SimulaMet}
Team SimulaMet are Nikhil Kumar Tomar, Debesh Jha, Michael A. Riegler and Pal Halvorsen from SimulaMet, UiT The Arctic University of Norway and Oslo Metropolitan University. They participated in all three tasks of the sub-challenge and used the same approach for all three tasks. The SimulaMet architecture was inspired by encoder-decoder networks such as ResUNet++ \cite{jha2019resunet++} and the success of pre-trained networks for medical image segmentation tasks \cite{alam2020automatic}.
A block diagram of the proposed architecture can be found in Figure \ref{fig:models-participants-1}-b1. As depicted in the figure, the proposed architecture begins with a pre-trained ResNet50 \cite{he2016deep} block, acting as the encoder for the architecture, where the images are passed into the encoder generating feature maps. These feature maps are then passed through various subsequent decoder blocks. The details describing the decoder blocks can be found in Figure \ref{fig:models-participants-1}-b2. Each decoder block begins with a bi-linear upsampling, which doubles the feature maps’ spatial dimensions. After this, the upsampled feature maps are passed through a 1 $\times$ 1 convolution, followed by a batch normalization and a Rectified Linear Unit (ReLU) activation, as seen in Figure \ref{fig:models-participants-1}-b2. Next, the output of the 1 $\times$ 1 convolution is concatenated with the appropriate feature map from the ResNet50 encoder through the skip connections. The skip connections help to carry forward the encoder layer feature map to the decoder layer, which helps the decoder to generate better semantic features. Next, the concatenated feature map is passed through a residual block (see Figure \ref{fig:models-participants-1}-b2 after concatenation operation). The residual block consists of two 3 $\times$ 3 convolutions and an identity mapping connecting the input and output of the convolution layer. Here, each convolution layer is followed by a batch normalization and ReLU activation. The output of the last decoder block is passed through a 1 $\times$ 1 convolution and finally through a softmax activation function to generate the segmentation mask. To implement the architecture, the TensorFlow framework with Keras API was used. For all three tasks, a ResNet50 backbone was used and the total number of parameters for the model was 9.28 M. The hyperparameters used for the models submitted to each sub-task were: learning rate of 1$e^{-4}$, batch size of 16, using the Adam optimizer and a cross-entropy loss. The models were trained using an NVIDIA V100 Tensor Core GPU. Data augmentation was applied during training of the models for all sub-tasks. They only used the image samples from the Cataract Dataset for Image Segmentation (CaDIS) while training the model. The original size of the images is 540$\times$960. The dataset consists of frames from 25 different videos, out of which, they used the frames of 22 videos for training and 3 videos for validation. The training dataset was augmented and resized using center crop, horizontal flip, vertical flip, grayscale, and cutout. The validation dataset is not augmented and is directly resized into 270$\times$480.
\subsection{SRV-WEISS}
Team SRV-WEISS are Sophia Bano and Uddhav Vaghela from Wellcome/EPSRC Centre for Interventional and Surgical Sciences (WEISS), University College London and Cambridge University NHS Trust, UK. For this challenge, they first performed an ablation study for model selection through prototype experiments in which only a subset of the provided training data was used. They experimented with both DeepLabV3 \cite{chen2017rethinking} and DeeplabV3+ \cite{chen2018encoder} architectures and Mobilenet-v2 \cite{sandler2018mobilenetv2}, ResneXt \cite{xie2017aggregated}, EfficientNet \cite{tan2019efficientnet} and Dual path network (DPN) \cite{chen2017dual} encoders since these segmentation architectures and encoders were shown to be effective and efficient in the Computer Vision literature. They found the combination of DeepLabV3 \cite{chen2017rethinking} with DPN-68 \cite{tan2019efficientnet} to be the most promising one in terms of the overall network performance. DeepLabv3 \cite{chen2017rethinking} uses ASPP that enables encoding of multi-scale contextual information while DPN \cite{tan2019efficientnet} shares common features just like ResNet alongside maintaining the flexibility to explore new features like DenseNet. For training, they used all provided images from the 25 videos in original resolution and applied a random crop of 320$\times$ 320 pixel resolution with random flipping, rotation of $\pm$30 degrees, brightness and contrast variation and image blurring with a 3$\times$3 kernel. An initial learning rate of 0.001 was used with step decay at every 50$_{th}$ epoch until 150 epochs and with an Adam optimizer. The network was trained for 200 epochs with early stopping based on the criterion to terminate the training when there was no improvement in the training accuracy for 20 epochs. Binary cross-entropy loss was used for Tasks I and II and weighted binary cross-entropy was used for task III to handle class imbalance. For testing, they performed test-time data augmentation (for the Task I only) by applying flipping, rotation and intensity change during testing and found an improvement of 0.5\% in the mIoU values.

\begin{figure*}[t!]
\includegraphics[width=\textwidth]{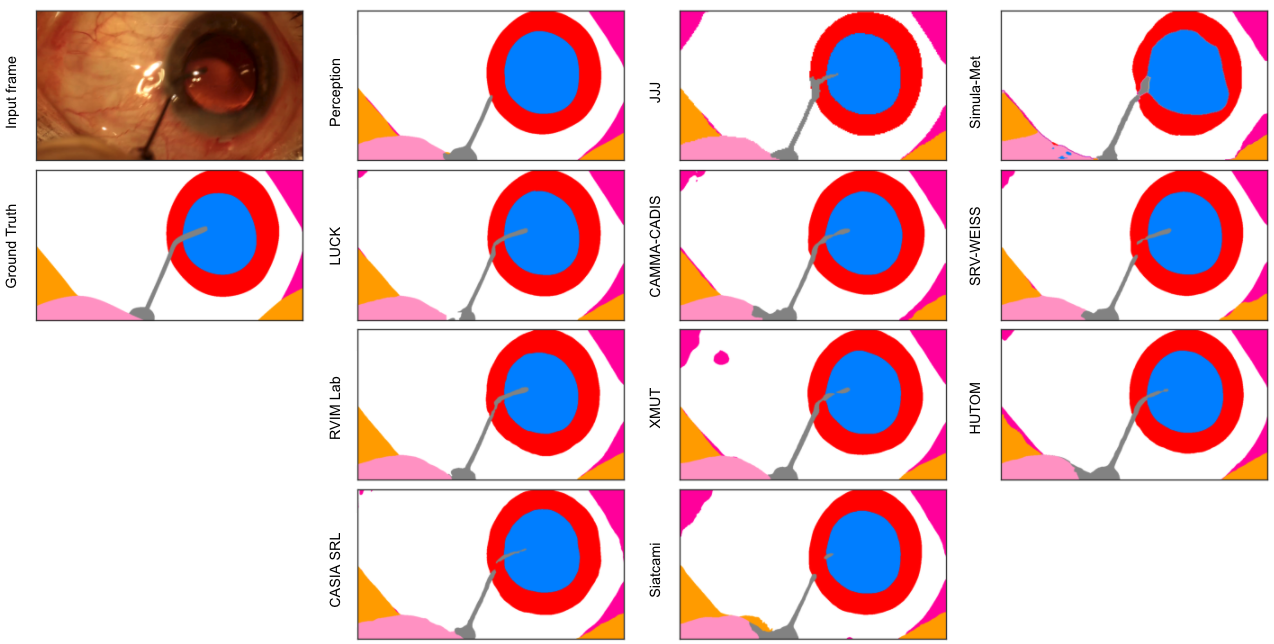}
\caption{Example frames with reference segmentation mask and model predictions for all teams for Task I. (Colormap: \textcolor{pupil}{\rule{0.5cm}{0.25cm}} Pupil, \textcolor{iris}{\rule{0.5cm}{0.25cm}} Iris,  $^{\framebox{ \textcolor{cornea}{\rule{0.3cm}{0.001cm}}}}$ Cornea, \textcolor{skin}{\rule{0.5cm}{0.25cm}} Skin, \textcolor{tape}{\rule{0.5cm}{0.25cm}} Surgical tape, \textcolor{retractors}{\rule{0.5cm}{0.25cm}} Eye retractors and \textcolor{hcannula}{\rule{0.5cm}{0.25cm}} Instrument) }
\label{fig:results_exp1}
\end{figure*}

\begin{table*}[t!]
\caption{IoU per class per team for Task I}
\label{tab:per-class-iou-task1}
\setlength{\tabcolsep}{5pt}
\begin{tabular}{l|ccccccccccc}
 & XMUT & Perception & HUTOM & Siatcami & CASIA SRL & RVIM Lab & JJJ & SimulaMet & SRV-WEISS & \begin{tabular}[c]{@{}c@{}}CAMMA-\\ CADIS\end{tabular} & LUCK \\
 \hline
Pupil & 94.32 & 94.27 & 93.87 & 94.29 & 93.92 & 94.53 & 93.63 & 89.85 & \textbf{94.56} & 94.32 & \textbf{94.56} \\
Surgical Tape & 84.93 & 84.14 & 81.38 & 83.94 & 83.1 & 83.79 & 84.03 & 68.94 & \textbf{89.02} & 77.7 & 86.82 \\
Hand & 79.74 & 77.16 & 78.55 & 78.35 & 76.35 & 77.84 & 78.11 & 53.47 & \textbf{79.75} & 78.38 & 79.01 \\
Eye Retractors & 80.23 & 79.68 & \textbf{81.24} & 77.18 & 77.64 & 81.09 & 74.9 & 41.82 & 78.22 & 75.25 & 81.05 \\
Iris & 84.43 & 84.78 & 83.87 & 85.18 & 84 & 85.14 & 83.02 & 73.14 & \textbf{85.29} & 84.2 & 85.28 \\
Skin & 82.88 & 83.34 & 82.94 & 80.91 & 79.77 & 83.72 & 82.43 & 40.34 & \textbf{87.12} & 78.74 & 86.09 \\
Cornea & 93.94 & 94.38 & 94.71 & 93.26 & 93.18 & 94.63 & 93.46 & 81.74 & 94.83 & 93.83 & \textbf{94.96} \\
Instrument & 79.91 & 80.83 & 80.89 & 79.7 & 78.93 & \textbf{82.94} & 75.87 & 51.22 & 81.34 & 77.87 & 82.43
\end{tabular}
\end{table*}

\begin{table}[t!]
\caption{Mean Intersection over Union [\%], number of videos in which highest mIoU was achieved and number of classes in which highest mIoU was achieved per team for Task I}
\label{tab:rank-task1}
\begin{tabular}{ccccc}
\# & Team & mIoU & \# videos   win & \# labels   win \\
\hline
\cellcolor[HTML]{D9EAD3}1 & \cellcolor[HTML]{D9EAD3}LUCK & \cellcolor[HTML]{D9EAD3} 86.27 & \cellcolor[HTML]{D9EAD3}4 & \cellcolor[HTML]{D9EAD3}1 \\
\cellcolor[HTML]{FFF2CC}2 & \cellcolor[HTML]{FFF2CC}SRV-WEISS & \cellcolor[HTML]{FFF2CC} 86.26 & \cellcolor[HTML]{FFF2CC}1 & \cellcolor[HTML]{FFF2CC}5 \\
\cellcolor[HTML]{FCE5CD}3 & \cellcolor[HTML]{FCE5CD}RVIM Lab & \cellcolor[HTML]{FCE5CD} 85.46 & \cellcolor[HTML]{FCE5CD}5 & \cellcolor[HTML]{FCE5CD}1 \\
4 & XMUT & 85.05 & - & - \\
5 & Perception & 84.82 & - & - \\
6 & HUTOM & 84.68 & - & 1 \\
7 & Siatcami & 84.1 & - & - \\
8 & CASIA SRL & 83.36 & - & - \\
9 & JJJ & 83.18 & - & - \\
10 & CAMMA-CADIS & 82.54 & - & - \\
11 & SimulaMet & 62.56 & - & -
\end{tabular}
\end{table}

\begin{table}[t!]
\caption{Rank stability for Task I}
\label{tab:rank-stability-task-1}
\setlength{\tabcolsep}{3pt}
\begin{tabular}{l|cccccc}
& \begin{tabular}[c]{@{}c@{}}Task I\\Rank\end{tabular} & Subset 1 & Subset 2 & Subset 3 & Subset 4 & \begin{tabular}[c]{@{}c@{}}Subset\\Mean\end{tabular} \\
\hline
Perception & 5 & 5 & \cellcolor[HTML]{D9EAD3}4 & \cellcolor[HTML]{D9EAD3}4 & 5 & \cellcolor[HTML]{D9EAD3}4 \\
JJJ & 9 & 9 & 9 & 9 & 9 & 9 \\
SimulaMet & 11 & 11 & 11 & 11 & 11 & 11 \\
LUCK & 1 & \cellcolor[HTML]{F4CCCC}2 & \cellcolor[HTML]{F4CCCC}2 & 1 & 1 & \cellcolor[HTML]{F4CCCC}2 \\
\begin{tabular}[l]{@{}l@{}}CAMMA-\\ CADIS\end{tabular} & 10 & 10 & 10 & 10 & 10 & 10 \\
SRV-WEISS & 2 & \cellcolor[HTML]{D9EAD3}1 & \cellcolor[HTML]{D9EAD3}1 & 2 & 2 & \cellcolor[HTML]{D9EAD3}1 \\
RVIM Lab & 3 & 3 & 3 & 3 & 3 & 3 \\
XMUT & 4 & 4 & \cellcolor[HTML]{F4CCCC}5 & \cellcolor[HTML]{F4CCCC}5 & 4 & \cellcolor[HTML]{F4CCCC}5 \\
HUTOM & 6 & 6 & 6 & 6 & 6 & 6 \\
CASIA SRL & 8 & 8 & 8 & 8 & 8 & 8 \\
Siatcami & 7 & 7 & 7 & 7 & 7 & 7
\end{tabular}
\end{table}

\subsection{XMUT}
Team XMUT are Jianyuan Hong , Haili Ye, Feihong Huang and Da-Han Wang from the School of Computer and Information Engineering, Xiamen University of Technology, China. Their method was based on DaNet \cite{fu2019dual} and ResNetV1. The main contribution of the DANet+ResNetV1 solution was to build a model based on the semantic segmentation method of DANet. The model can effectively aggregate the context information of the image, and through data enhancement and loss design, further improves the model's segmentation precision of objects in the image. From the videos provided for training, 22 videos were selected as the train set and 3 videos were used as the validation set. Resnet50 used images with a size of 540$\times$960 pixels. The data was standardized in the pre-processing stage. In terms of data augmentation, this solution only performed translation and flip operations on part of the data. The network used the cross-entropy loss, Lovasz loss \cite{berman2018lovasz}, mIoU hybrid loss and SGD optimizer for training. In the training phase, all parameters of the network were initialized from the pre-trained ImageNet model. After up to 30 epochs, the optimal batch size was found to be 6 and the momentum to be 0.9.

\section{Assessment methods}
The metric that was used to assess model performance was the mean Intersection over Union (mIoU) as this is the most representative metric assessing segmentation tasks. The formulation for the mIoU metric is defined as follows:
\begin{align}
    \text{mIoU} &= \frac{1}{N} \sum_{i=1}^{N} \frac{p_{ii}}{\sum_{j=1}^{N} p_{ij} - p_{ii} + \sum_{j=1}^{N} p_{ji}}, \ \ i,j =1,..,N
\end{align}
were $N$ the number of classes and $p_{ij}$ the number of pixels predicted as class $i$ and labelled as class $j$.

\begin{figure*}{t!}
\includegraphics[width=\textwidth]{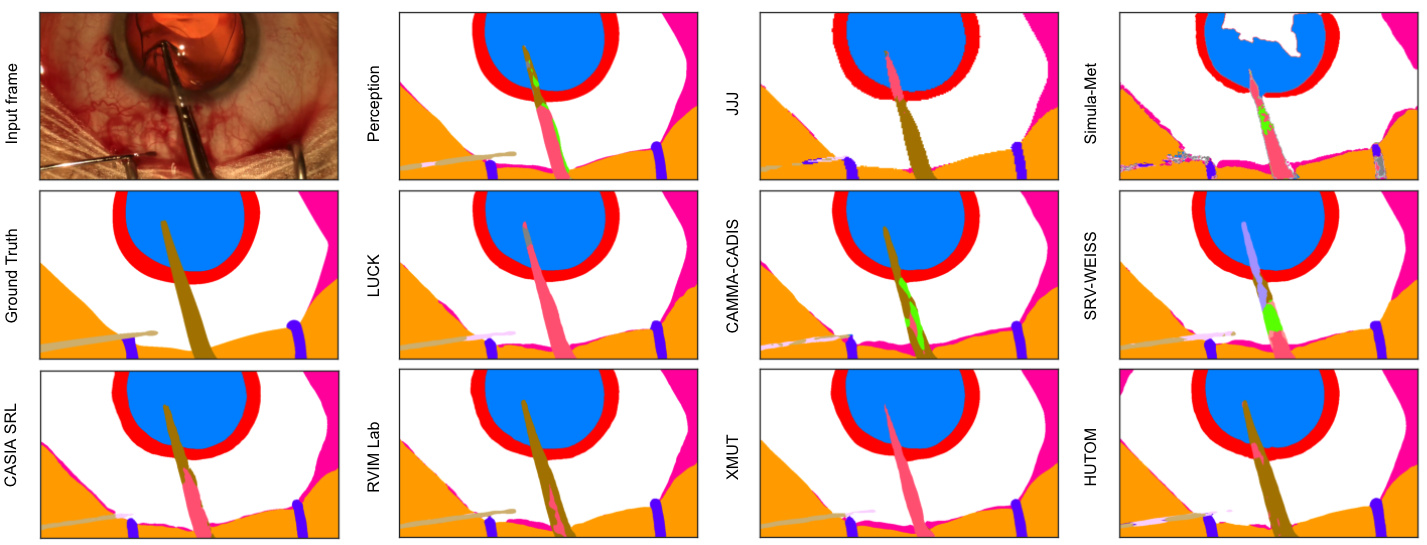}
\caption{Example frames with reference segmentation mask and model predictions for all teams for Task II. (Colormap: \textcolor{pupil}{\rule{0.5cm}{0.25cm}} Pupil, \textcolor{iris}{\rule{0.5cm}{0.25cm}} Iris,  $^{\framebox{ \textcolor{cornea}{\rule{0.3cm}{0.001cm}}}}$ Cornea, \textcolor{skin}{\rule{0.5cm}{0.25cm}} Skin, \textcolor{tape}{\rule{0.5cm}{0.25cm}} Surgical tape, \textcolor{retractors}{\rule{0.5cm}{0.25cm}} Eye retractors, \textcolor{micromanipulator}{\rule{0.5cm}{0.25cm}} Micromanipulator, \textcolor{capforceps}{\rule{0.5cm}{0.25cm}} Capsulorhexis forceps,  \textcolor{bonn}{\rule{0.5cm}{0.25cm}} Bonn forceps,  \textcolor{phaco}{\rule{0.5cm}{0.25cm}} Phacoemulsification handpiece, \textcolor{capcyst}{\rule{0.5cm}{0.25cm}} Capsulorhexis cystotome) }
\label{fig:results_exp2}
\end{figure*}
\begin{table*}[t!]
\caption{IoU per class per team for Task II}
\label{tab:per-class-iou-task2}
\begin{tabular}{l|cccccccccc}
\rowcolor[HTML]{FFFFFF} 
\multicolumn{1}{c|}{\cellcolor[HTML]{FFFFFF}{\color[HTML]{333333} \textbf{}}} & XMUT & Perception & HUTOM & CASIA SRL & RVIM Lab & JJJ & SimulaMet & SRV-WEISS & \begin{tabular}[c]{@{}c@{}}CAMMA-\\ CADIS\end{tabular} & LUCK \\
\hline
\cellcolor[HTML]{FFFFFF}{\color[HTML]{333333} Pupil} & 93.97 & 94.23 & 93.54 & 93.86 & 94.23 & 93.1 & 91.49 & 94.38 & 94.24 & \textbf{94.54} \\
\cellcolor[HTML]{FFFFFF}{\color[HTML]{333333} Surgical Tape} & 83.72 & 80.1 & 80.18 & 78.13 & 83.12 & \textbf{85.59} & 76.39 & 81.42 & 77.53 & 85.24 \\
\cellcolor[HTML]{FFFFFF}{\color[HTML]{333333} Hand} & 79.54 & 78.7 & 78.73 & 78.04 & 78.41 & \textbf{79.59} & 56.38 & 78.8 & 78.22 & 79.08 \\
\cellcolor[HTML]{FFFFFF}{\color[HTML]{333333} Eye Retractors} & 79.1 & 78.78 & 80.84 & 77.52 & 80.6 & 75.41 & 43.1 & 77.21 & 75.89 & \textbf{81.06} \\
\cellcolor[HTML]{FFFFFF}{\color[HTML]{333333} Iris} & 84.09 & 84.32 & 83.28 & 84.1 & 84.61 & 82.97 & 74.17 & 84.76 & 84.12 & \textbf{85.28} \\
\cellcolor[HTML]{FFFFFF}{\color[HTML]{333333} Skin} & 81.79 & 80.93 & 81.94 & 77.99 & 83.52 & 83.06 & 57.38 & 83.13 & 78.79 & \textbf{85.06} \\
\cellcolor[HTML]{FFFFFF}{\color[HTML]{333333} Cornea} & 93.86 & 94.24 & 94.49 & 93.37 & 94.61 & 93.47 & 86.29 & 94.69 & 93.86 & \textbf{94.94} \\
\cellcolor[HTML]{FFFFFF}{\color[HTML]{333333} Cannula} & 73.08 & 69.23 & 71.73 & 68.93 & \textbf{74.61} & 63.31 & 28.85 & 67.92 & 67.98 & 70.63 \\
Cap.  Cystotome & 77.86 & 71.45 & 80.93 & 78.75 & \textbf{83.26} & 70.81 & 4.04 & 60.3 & 70.91 & 82.52 \\
Tissue Forceps & 74.92 & 77.91 & 79.32 & 80.54 & \textbf{85.93} & 74.79 & 34.61 & 79.46 & 70.29 & 82.63 \\
Primary  Knife & 89.22 & 89.37 & 88.31 & 88.69 & \textbf{93.31} & 84.31 & 50.3 & 88.3 & 89.82 & 91.43 \\
Ph. Handpiece & 78.53 & 77.52 & 85.12 & 80.56 & 85.88 & 81.34 & 40.17 & 81.92 & 80.85 & \textbf{85.9} \\
Lens  Injector & 74.51 & 74.38 & \textbf{79.34} & 70.87 & 77.67 & 70.28 & 35.53 & 79.35 & 70.25 & 78.52 \\
I/A  Handpiece & 81.38 & 81.03 & 84.57 & 81.32 & \textbf{86.1} & 80.76 & 56.9 & 83.47 & 84.94 & 83.8 \\
Secondary  Knife & 86.4 & 85.54 & 86.61 & 82.81 & \textbf{87.67} & 80.48 & 39.14 & 83.34 & 84.52 & 83.22 \\
Micromanipulator & 63.96 & 61.86 & 67.24 & 61.52 & \textbf{68.9} & 55.02 & 1.26 & 60.3 & 63.06 & 61.84 \\
Cap.  Forceps & 45.52 & 45.68 & 82.7 & 62.95 & \textbf{83.02} & 74.5 & 0 & 17.2 & 65.78 & 52.91
\end{tabular}
\end{table*}

\begin{table}[t!]
\caption{Mean Intersection over Union [\%], number of videos in which highest mIoU was achieved and number of classes in which highest mIoU was achieved per team for Task II}
\label{tab:rank-task2}
\begin{tabular}{ccccc}
\# & Team & mIoU & \# videos win & \# labels win \\
\hline
\cellcolor[HTML]{D9EAD3}1 & \cellcolor[HTML]{D9EAD3}RVIM Lab & \cellcolor[HTML]{D9EAD3} 83.85 & \cellcolor[HTML]{D9EAD3}6 & \cellcolor[HTML]{D9EAD3}8 \\
\cellcolor[HTML]{FFF2CC}2 & \cellcolor[HTML]{FFF2CC}HUTOM & \cellcolor[HTML]{FFF2CC} 82.29 & \cellcolor[HTML]{FFF2CC}- & \cellcolor[HTML]{FFF2CC}- \\
\cellcolor[HTML]{FCE5CD}3 & \cellcolor[HTML]{FCE5CD}LUCK & \cellcolor[HTML]{FCE5CD} 81.09 & \cellcolor[HTML]{FCE5CD}1 & \cellcolor[HTML]{FCE5CD}6 \\
4 & XMUT & 78.91 & 1 & - \\
5 & CASIA SRL & 78.82 & - & - \\
6 & CAMMA-CADIS & 78.3 & 1 & - \\
7 & JJJ & 78.16 & - & 2 \\
8 & Perception & 77.96 & - & - \\
9 & SRV-WEISS & 76.23 & 1 & 1 \\
10 & SimulaMet & 45.65 & - & - \\
11 & Siatcami & - & - & -
\end{tabular}
\end{table}

\begin{table}[t!]
\caption{Rank stability for Task II}
\label{tab:rank-stability-task-2}
\setlength{\tabcolsep}{3pt}
\begin{tabular}{l|cccccc}
 & \multicolumn{1}{l}{\begin{tabular}[c]{@{}l@{}}Task II\\ Rank\end{tabular}} & \multicolumn{1}{l}{Subset 1} & \multicolumn{1}{l}{Subset 2} & \multicolumn{1}{l}{Subset 3} & \multicolumn{1}{l}{Subset 4} & \multicolumn{1}{l}{\begin{tabular}[c]{@{}l@{}}Subset \\ Mean\end{tabular}} \\
 \hline
Perception & 8 & \cellcolor[HTML]{D9EAD3}6 & \cellcolor[HTML]{D9EAD3}6 & \cellcolor[HTML]{D9EAD3}7 & \cellcolor[HTML]{D9EAD3}7 & \cellcolor[HTML]{D9EAD3}6 \\
JJJ & 7 & \cellcolor[HTML]{F4CCCC}8 & 7 & \cellcolor[HTML]{F4CCCC}8 & \cellcolor[HTML]{F4CCCC}8 & \cellcolor[HTML]{F4CCCC}8 \\
SimulaMet & 10 & 10 & 10 & 10 & 10 & 10 \\
LUCK & 3 & 3 & 3 & 3 & 3 & 3 \\
\begin{tabular}[c]{@{}l@{}}CAMMA-\\ CADIS\end{tabular} & 6 & \cellcolor[HTML]{F4CCCC}7 & \cellcolor[HTML]{F4CCCC}8 & \cellcolor[HTML]{D9EAD3}5 & \cellcolor[HTML]{D9EAD3}5 & \cellcolor[HTML]{F4CCCC}7 \\
SRV-WEISS & 9 & 9 & 9 & 9 & 9 & 9 \\
RVIM Lab & 1 & 1 & 1 & 1 & 1 & 1 \\
XMUT & 4 & 4 & 4 & 4 & 4 & 4 \\
HUTOM & 2 & 2 & 2 & 2 & 2 & 2 \\
CASIA SRL & 5 & 5 & 5 & \cellcolor[HTML]{F4CCCC}6 & \cellcolor[HTML]{F4CCCC}6 & 5 \\
Siatcami & \cellcolor[HTML]{EFEFEF}- & \cellcolor[HTML]{EFEFEF}- & \cellcolor[HTML]{EFEFEF}- & \cellcolor[HTML]{EFEFEF}- & \cellcolor[HTML]{EFEFEF}- & \cellcolor[HTML]{EFEFEF}-
\end{tabular}
\end{table}

\begin{figure*}[t!]
\includegraphics[width=\textwidth]{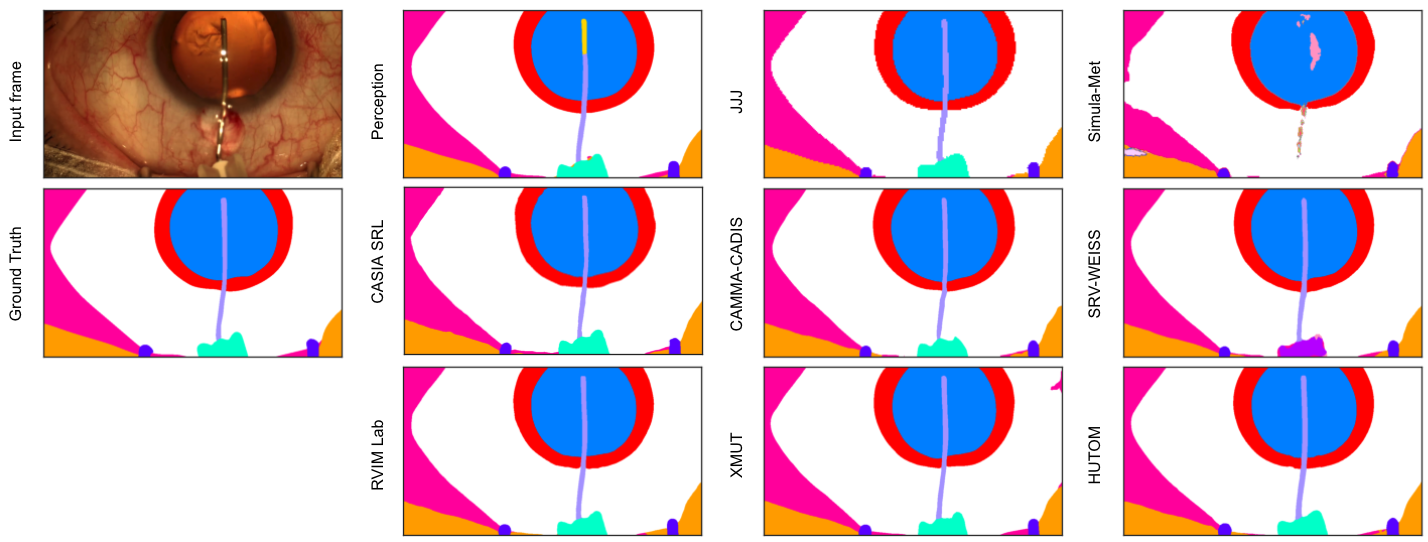}
\caption{Example frames with reference segmentation mask and model predictions for all teams for Task III. (Colormap: \textcolor{pupil}{\rule{0.5cm}{0.25cm}} Pupil, \textcolor{iris}{\rule{0.5cm}{0.25cm}} Iris,  $^{\framebox{ \textcolor{cornea}{\rule{0.3cm}{0.001cm}}}}$ Cornea, \textcolor{skin}{\rule{0.5cm}{0.25cm}} Skin, \textcolor{tape}{\rule{0.5cm}{0.25cm}} Surgical tape, \textcolor{retractors}{\rule{0.5cm}{0.25cm}} Eye retractors, \textcolor{capcyst}{\rule{0.5cm}{0.25cm}} Capsulorhexis cystotome, \textcolor{vcannula}{\rule{0.5cm}{0.25cm}} Viscoelastic cannula, \textcolor{secondknifehandle}{\rule{0.5cm}{0.25cm}} Secondary knife handle, \textcolor{hand}{\rule{0.5cm}{0.25cm}} Hand)}
\label{fig:results_exp3}
\end{figure*}
\begin{table}[t!]
\caption{Mean Intersection over Union, Number of videos where highest mIoU was achieved and number of classes where highest mIoU was achieved per team for Task III}
\label{tab:rank-task3}
\begin{tabular}{ccccc}
\# & Team & mIoU & \# videos win & \# labels win \\
\hline
\cellcolor[HTML]{D9EAD3}1 & \cellcolor[HTML]{D9EAD3}RVIM Lab & \cellcolor[HTML]{D9EAD3}0.7793 & \cellcolor[HTML]{D9EAD3}5 & \cellcolor[HTML]{D9EAD3}10 \\
\cellcolor[HTML]{FFF2CC}2 & \cellcolor[HTML]{FFF2CC}HUTOM & \cellcolor[HTML]{FFF2CC}0.7739 & \cellcolor[HTML]{FFF2CC}1 & \cellcolor[HTML]{FFF2CC}9 \\
\cellcolor[HTML]{FCE5CD}3 & \cellcolor[HTML]{FCE5CD}XMUT & \cellcolor[HTML]{FCE5CD}0.765 & \cellcolor[HTML]{FCE5CD}3 & \cellcolor[HTML]{FCE5CD}1 \\
4 & CAMMA-CADIS & 0.7537 & 1 & - \\
5 & CASIA SRL & 0.7278 & - & - \\
6 & Perception & 0.7122 & - & - \\
7 & JJJ & 0.7098 & - & - \\
8 & SRV-WEISS & 0.6407 & - & 4 \\
9 & SimulaMet & 0.3153 & - & - \\
10 & LUCK & - & - & - \\
11 & Siatcami & - & - & -
\end{tabular}
\end{table}

\section{Results}
\subsection{Task I}
For Task I, 10 out of 11 teams achieved a mIoU over 0.825. The three teams that had the highest mIoUs were LUCK, SRV-WEISS and RVIM Lab with 0.8627, 0.8626 and 0.8546 respectively (Table \ref{tab:rank-task1}). In this task, all instrument types are merged into one class making the distribution among the classes more balanced (Table \ref{fig:task-dist} - first column). Therefore, as this tasks is focusing on segmentation of 8 classes which are very distinct among each other, the failure cases for the models occur mostly from missed instrument parts. This is shown in the examples shown in Figure \ref{fig:results_exp1}. The IoUs per class for Task I are also given in Table \ref{tab:per-class-iou-task1}. It can be seen that all teams achieve relatively high per class IoUs.
\begin{table}[t!]
\caption{Rank stability for Task III}
\label{tab:rank-stability-task-3}
\setlength{\tabcolsep}{3pt}
\begin{tabular}{l|cccccc}
 & \multicolumn{1}{l}{\begin{tabular}[c]{@{}l@{}}Task III\\ Rank\end{tabular}} & \multicolumn{1}{l}{Subset 1} & \multicolumn{1}{l}{Subset 2} & \multicolumn{1}{l}{Subset 3} & \multicolumn{1}{l}{Subset 4} & \multicolumn{1}{l}{\begin{tabular}[c]{@{}l@{}}Subset \\ Mean\end{tabular}} \\
 \hline
Perception & 6 & \cellcolor[HTML]{F4CCCC}7 & \cellcolor[HTML]{D9EAD3}5 & 6 & \cellcolor[HTML]{F4CCCC}7 & \cellcolor[HTML]{F4CCCC}7 \\
JJJ & 7 & \cellcolor[HTML]{D9EAD3}6 & 7 & 7 & \cellcolor[HTML]{D9EAD3}6 & \cellcolor[HTML]{D9EAD3}6 \\
SimulaMet & 9 & 9 & 9 & 9 & 9 & 9 \\
LUCK & \cellcolor[HTML]{EFEFEF}- & \cellcolor[HTML]{EFEFEF}- & \cellcolor[HTML]{EFEFEF}- & \cellcolor[HTML]{EFEFEF}- & \cellcolor[HTML]{EFEFEF}- & \cellcolor[HTML]{EFEFEF}- \\
\begin{tabular}[c]{@{}l@{}}CAMMA-\\ CADIS\end{tabular} & 4 & \cellcolor[HTML]{F4CCCC}5 & 4 & \cellcolor[HTML]{D9EAD3}3 & 4 & 4 \\
SRV-WEISS & 8 & 8 & 8 & 8 & 8 & 8 \\
RVIM Lab & 1 & 1 & 1 & 1 & 1 & 1 \\
XMUT & 3 & 3 & 3 & \cellcolor[HTML]{F4CCCC}4 & 3 & 3 \\
HUTOM & 2 & 2 & 2 & 2 & 2 & 2 \\
CASIA SRL & 5 & \cellcolor[HTML]{D9EAD3}4 & \cellcolor[HTML]{F4CCCC}6 & 5 & 5 & 5 \\
Siatcami & \cellcolor[HTML]{EFEFEF}- & \cellcolor[HTML]{EFEFEF}- & \cellcolor[HTML]{EFEFEF}- & \cellcolor[HTML]{EFEFEF}- & \cellcolor[HTML]{EFEFEF}- & \cellcolor[HTML]{EFEFEF}-
\end{tabular}
\end{table}

\begin{table*}[t!]
\caption{IoU per class per team for Task III}
\label{tab:per-class-iou-task3}
\begin{tabular}{l|rrrrrrrrr}
 & \multicolumn{1}{l}{XMUT} & \multicolumn{1}{l}{Perception} & \multicolumn{1}{l}{HUTOM} & \multicolumn{1}{l}{CASIA SRL} & \multicolumn{1}{l}{RVIM Lab} & \multicolumn{1}{l}{JJJ} & \multicolumn{1}{l}{SimulaMet} & \multicolumn{1}{l}{SRV-WEISS} & \multicolumn{1}{l}{CAMMA-CADIS} \\
 \hline
Pupil & 93.67 & 94.37 & 94.1 & 93.81 & 94.13 & 93.37 & 90.56 & \textbf{94.56} & 94.36 \\
Surgical Tape & 83.69 & 83.15 & 80.39 & 78.97 & 83.28 & 83.66 & 77.86 & 82.12 & 77.86 \\
Hand & 79.18 & 76.78 & 77.23 & 78.1 & \textbf{79.42} & 78.48 & 51.98 & 78.52 & 77.9 \\
Eye Retractors & 77.97 & 79.15 & \textbf{81.83} & 78 & 80.08 & 75.54 & 47.53 & 77.34 & 73.96 \\
Iris & 83.92 & 84.34 & 84.39 & 83.89 & 84.6 & 83.06 & 77.57 & \textbf{85.15} & 83.83 \\
Skin & 81.66 & 83.07 & 82.3 & 78.54 & 83.44 & 81.83 & 54.5 & \textbf{83.45} & 78.74 \\
Cornea & 93.68 & 94.32 & 94.78 & 93.36 & 94.52 & 93.43 & 85.67 & \textbf{94.71} & 93.66 \\
Hydro. Cannula & 76.57 & 72.14 & 81.15 & 68.24 & \textbf{79.45} & 60.74 & 4.73 & 66.85 & 71.51 \\
Visc. Cannula & 63.74 & 60.05 & 66.26 & 58.15 & \textbf{69.4} & 51.19 & 1.34 & 60.36 & 60.27 \\
Cap. Cystotome & 76.04 & 63.69 & 78.99 & 75.66 & \textbf{81.25} & 65.84 & 0.45 & 65.67 & 68.96 \\
Rycroft Cannula & 61.4 & 53.38 & \textbf{63.58} & 57.83 & 63.12 & 54.46 & 0.08 & 47.54 & 60.22 \\
Bonn Forceps & 70.41 & 76.83 & 80.69 & 81.97 & \textbf{83.26} & 75.37 & 28.75 & 81.33 & 70.58 \\
Primary Knife & 90.09 & 90.27 & 90.36 & 83.9 & \textbf{91.33} & 81.2 & 50.42 & 89.16 & 89.54 \\
Ph. Handpiece & 73.74 & 79.27 & \textbf{85.02} & 78.1 & 83.87 & 77.61 & 39.49 & 80.3 & 78.6 \\
Lens Injector & 69.53 & 74.02 & 76.71 & 71.16 & 76.53 & 71.31 & 39.02 & \textbf{79.55} & 69.55 \\
I/A Handpiece & 79.65 & 80.85 & 84.57 & 81.88 & \textbf{84.29} & 79.47 & 54.11 & 80.7 & 82.98 \\
Secondary Knife & 83.18 & 79.52 & 85.24 & 78.93 & \textbf{85.98} & 76.83 & 25.96 & 84.18 & 84.34 \\
Micromanipulator & 62.42 & 61.67 & \textbf{68.63} & 61.57 & 67.81 & 54.51 & 18.17 & 55.02 & 61.31 \\
I/A Handpiece Handle & 79.06 & 73.35 & 80.85 & 75.7 & \textbf{81.44} & 78.58 & 0.87 & 73.31 & 76.58 \\
Cap. Forceps & 45.91 & 23.01 & \textbf{86.18} & 69.68 & 81.55 & 68.35 & 0.23 & 0 & 65.04 \\
R. Cannula Handle & 41.6 & 41.67 & 39.22 & 37.41 & \textbf{46.12} & 33.45 & 0 & 0 & 31.11 \\
Ph. Handpiece Handle & 89.65 & \textbf{90.72} & 91.62 & 87.66 & 89.67 & 88.99 & 39 & 86.28 & 89.72 \\
Cap. Cystotome Handle & 90.15 & 86.67 & 91.21 & 88.07 & \textbf{92.09} & 86.12 & 0 & 0 & 84.45 \\
Sec. Knife Handle & 89.03 & 78.32 & 89.5 & 79.01 & \textbf{91.5} & 81.16 & 0 & 55.7 & 83.86
\end{tabular}
\end{table*}
\subsection{Task II}
In this task, 9 out of 10 participating teams achieved mIoUs between 0.7623 to 0.8385. As the number of classes are increased in this task compared to Task I, the imbalance issue is more evident, hence there is more variation between the participating solutions. It is worth noting that model ensembles demonstrated a benefit towards instrument segmentation, something which can be seen in the per class IoUs for Task II of Table \ref{tab:per-class-iou-task2}. Example outputs from all teams are shown in Figure \ref{fig:results_exp2}.

\subsection{Task III}
In Task III, class imbalanced is even larger with 25 classes to be segmented. The impact of it can be seen by comparing the mIoUS of Tables \ref{tab:rank-task2} and \ref{tab:rank-task3} and the per class IoUs (Tables \ref{tab:per-class-iou-task2},\ref{tab:per-class-iou-task3}). Similarly to Task II, model ensembles, test time augmentation and loss function that are more suited towards imbalanced datasets seemed to benefit multiple class segmentation. An example output from all participating teams for Task III can be seen in Figure \ref{fig:results_exp3}. 

\subsection{Rank stability}
In order to investigate the robustness of using the mIoU on the whole test set as a fair metric for model assessment, a rank stability test was performed. To this end, four subsets of the test set were created including 400 random images each. Then the mIoU for each participating solution for each task was calculated for these four subsets. From Tables \ref{tab:rank-stability-task-1}, \ref{tab:rank-stability-task-2} and \ref{tab:rank-stability-task-3}, it can be seen that in general the ranking is stable, with the exception of Task I for the two top ranked models whose performance was comparable (Table \ref{tab:rank-stability-task-1}).

\section{Conclusions}
Surgical computer vision has the potential to advance intra-operative computer-assisted surgical interventions and post-operative analysis of the surgical procedures. Deep learning algorithms could be used to benefit such applications. However, large training datasets from surgical videos are required to do so. To this end, by making labelled datasets public to the research community and assessing their solutions provides insights towards the advancement of the surgical computer vision field. In the 2020 CATARACTS Semantic Segmentation Challenge, which was part of the  2020 MICCAI EndoVis Challenge, the participating teams were assessed on anatomy and instrument segmentation in cataract surgery videos. Overall, all participating teams presented interesting approaches to the challenge's sub-tasks. It was seen that as the number of classes increases, class imbalance becomes more evident and presents the main problem to tackle in the challenge's dataset. Task I shows that when imbalance is mitigated by instrument grouping, the inherent differences of the proposed solutions are less evident. It was also seen that for Task II and III, model ensembles showed benefits towards multiple class instrument segmentation. Additionally, test time augmentation and losses that are more suitable for imbalanced datasets seemed to benefit multiple class segmentation. It is also worth noting that the public training and hidden test set had similar class distributions. In conclusion, while all solutions performed very well at distinguishing between anatomy and instruments, correctly identifying and achieving spatially consistent instrument segmentation is still an open problem.
\bibliographystyle{IEEEtran}
\bibliography{biblio}

\begin{thebibliography}{10}
\providecommand{\url}[1]{#1}
\csname url@samestyle\endcsname
\providecommand{\newblock}{\relax}
\providecommand{\bibinfo}[2]{#2}
\providecommand{\BIBentrySTDinterwordspacing}{\spaceskip=0pt\relax}
\providecommand{\BIBentryALTinterwordstretchfactor}{4}
\providecommand{\BIBentryALTinterwordspacing}{\spaceskip=\fontdimen2\font plus
\BIBentryALTinterwordstretchfactor\fontdimen3\font minus
  \fontdimen4\font\relax}
\providecommand{\BIBforeignlanguage}[2]{{%
\expandafter\ifx\csname l@#1\endcsname\relax
\typeout{** WARNING: IEEEtran.bst: No hyphenation pattern has been}%
\typeout{** loaded for the language `#1'. Using the pattern for}%
\typeout{** the default language instead.}%
\else
\language=\csname l@#1\endcsname
\fi
#2}}
\providecommand{\BIBdecl}{\relax}
\BIBdecl

\bibitem{cataracts}
H.~Al~Hajj \emph{et~al.}, ``Cataracts: Challenge on automatic tool annotation
  for cataract surgery,'' \emph{Medical image analysis}, vol.~52, pp. 24--41,
  2019.

\bibitem{endovis2017}
M.~Allan \emph{et~al.}, ``2017 {Robotic Instrument Segmentation Challenge},''
  \emph{arXiv preprint arXiv:1902.06426}, 2019.

\bibitem{allan20202018}
------, ``2018 {Robotic Scene Segmentation Challenge},'' \emph{arXiv preprint
  arXiv:2001.11190}, 2020.

\bibitem{grammatikopoulou2019cadis}
M.~Grammatikopoulou, E.~Flouty, A.~Kadkhodamohammadi, G.~Quellec, A.~Chow,
  J.~Nehme, I.~Luengo, and D.~Stoyanov, ``Cadis: Cataract dataset for image
  segmentation,'' \emph{arXiv preprint arXiv:1906.11586}, 2019.

\bibitem{ni2020pyramid}
Z.-L. Ni, G.-B. Bian, G.-A. Wang, X.-H. Zhou, Z.-G. Hou, H.-B. Chen, and X.-L.
  Xie, ``Pyramid attention aggregation network for semantic segmentation of
  surgical instruments,'' in \emph{Proceedings of the AAAI Conference on
  Artificial Intelligence}, vol.~34, no.~07, 2020, pp. 11\,782--11\,790.

\bibitem{zhu2019deformable}
X.~Zhu, H.~Hu, S.~Lin, and J.~Dai, ``Deformable convnets v2: More deformable,
  better results,'' in \emph{Proceedings of the IEEE/CVF Conference on Computer
  Vision and Pattern Recognition}, 2019, pp. 9308--9316.

\bibitem{zhang2020resnest}
H.~Zhang, C.~Wu, Z.~Zhang, Y.~Zhu, Z.~Zhang, H.~Lin, Y.~Sun, T.~He, J.~Mueller,
  R.~Manmatha \emph{et~al.}, ``Resnest: Split-attention networks,'' \emph{arXiv
  preprint arXiv:2004.08955}, 2020.

\bibitem{chen2018encoder}
L.-C. Chen, Y.~Zhu, G.~Papandreou, F.~Schroff, and H.~Adam, ``Encoder-decoder
  with atrous separable convolution for semantic image segmentation,'' in
  \emph{Proceedings of the European conference on computer vision (ECCV)},
  2018, pp. 801--818.

\bibitem{wang2020deep}
J.~Wang, K.~Sun, T.~Cheng, B.~Jiang, C.~Deng, Y.~Zhao, D.~Liu, Y.~Mu, M.~Tan,
  X.~Wang \emph{et~al.}, ``Deep high-resolution representation learning for
  visual recognition,'' \emph{IEEE transactions on pattern analysis and machine
  intelligence}, 2020.

\bibitem{islam2019learning}
M.~Islam, Y.~Li, and H.~Ren, ``Learning where to look while tracking
  instruments in robot-assisted surgery,'' in \emph{International Conference on
  Medical Image Computing and Computer-Assisted Intervention}.\hskip 1em plus
  0.5em minus 0.4em\relax Springer, 2019, pp. 412--420.

\bibitem{islam2021st}
M.~Islam, V.~Vibashan, C.~M. Lim, and H.~Ren, ``St-mtl: Spatio-temporal
  multitask learning model to predict scanpath while tracking instruments in
  robotic surgery,'' \emph{Medical Image Analysis}, vol.~67, p. 101837, 2021.

\bibitem{li2018adaptive}
Y.~Li, N.~Wang, J.~Shi, X.~Hou, and J.~Liu, ``Adaptive batch normalization for
  practical domain adaptation,'' \emph{Pattern Recognition}, vol.~80, pp.
  109--117, 2018.

\bibitem{yuan2019object}
Y.~Yuan, X.~Chen, and J.~Wang, ``Object-contextual representations for semantic
  segmentation,'' \emph{arXiv preprint arXiv:1909.11065}, 2019.

\bibitem{xiao2018unified}
T.~Xiao, Y.~Liu, B.~Zhou, Y.~Jiang, and J.~Sun, ``Unified perceptual parsing
  for scene understanding,'' in \emph{Proceedings of the European Conference on
  Computer Vision (ECCV)}, 2018, pp. 418--434.

\bibitem{berman2018lovasz}
M.~Berman, A.~R. Triki, and M.~B. Blaschko, ``The lov{\'a}sz-softmax loss: A
  tractable surrogate for the optimization of the intersection-over-union
  measure in neural networks,'' in \emph{Proceedings of the IEEE Conference on
  Computer Vision and Pattern Recognition}, 2018, pp. 4413--4421.

\bibitem{karpathy}
A.~Karpathy, ``A recipe for training neural networks,'' in
  \emph{http://karpathy.github.io/2019/04/25/recipe/}, 2019.

\bibitem{gupta2019lvis}
A.~Gupta, P.~Dollar, and R.~Girshick, ``Lvis: A dataset for large vocabulary
  instance segmentation,'' in \emph{Proceedings of the IEEE/CVF Conference on
  Computer Vision and Pattern Recognition}, 2019, pp. 5356--5364.

\bibitem{ouali2020semi}
Y.~Ouali, C.~Hudelot, and M.~Tami, ``Semi-supervised semantic segmentation with
  cross-consistency training,'' in \emph{Proceedings of the IEEE/CVF Conference
  on Computer Vision and Pattern Recognition}, 2020, pp. 12\,674--12\,684.

\bibitem{jha2019resunet++}
D.~Jha, P.~H. Smedsrud, M.~A. Riegler, D.~Johansen, T.~De~Lange, P.~Halvorsen,
  and H.~D. Johansen, ``Resunet++: An advanced architecture for medical image
  segmentation,'' in \emph{2019 IEEE International Symposium on Multimedia
  (ISM)}.\hskip 1em plus 0.5em minus 0.4em\relax IEEE, 2019, pp. 225--2255.

\bibitem{alam2020automatic}
S.~Alam, N.~K. Tomar, A.~Thakur, D.~Jha, and A.~Rauniyar, ``Automatic polyp
  segmentation using u-net-resnet50,'' \emph{arXiv preprint arXiv:2012.15247},
  2020.

\bibitem{he2016deep}
K.~He, X.~Zhang, S.~Ren, and J.~Sun, ``Deep residual learning for image
  recognition,'' in \emph{Proceedings of the IEEE conference on computer vision
  and pattern recognition}, 2016, pp. 770--778.

\bibitem{chen2017rethinking}
L.-C. Chen, G.~Papandreou, F.~Schroff, and H.~Adam, ``Rethinking atrous
  convolution for semantic image segmentation,'' \emph{arXiv preprint
  arXiv:1706.05587}, 2017.

\bibitem{sandler2018mobilenetv2}
M.~Sandler, A.~Howard, M.~Zhu, A.~Zhmoginov, and L.-C. Chen, ``Mobilenetv2:
  Inverted residuals and linear bottlenecks,'' in \emph{Proceedings of the IEEE
  conference on computer vision and pattern recognition}, 2018, pp. 4510--4520.

\bibitem{xie2017aggregated}
S.~Xie, R.~Girshick, P.~Doll{\'a}r, Z.~Tu, and K.~He, ``Aggregated residual
  transformations for deep neural networks,'' in \emph{Proceedings of the IEEE
  conference on computer vision and pattern recognition}, 2017, pp. 1492--1500.

\bibitem{tan2019efficientnet}
M.~Tan and Q.~Le, ``Efficientnet: Rethinking model scaling for convolutional
  neural networks,'' in \emph{International Conference on Machine
  Learning}.\hskip 1em plus 0.5em minus 0.4em\relax PMLR, 2019, pp. 6105--6114.

\bibitem{chen2017dual}
Y.~Chen, J.~Li, H.~Xiao, X.~Jin, S.~Yan, and J.~Feng, ``Dual path networks,''
  \emph{arXiv preprint arXiv:1707.01629}, 2017.

\bibitem{fu2019dual}
J.~Fu, J.~Liu, H.~Tian, Y.~Li, Y.~Bao, Z.~Fang, and H.~Lu, ``Dual attention
  network for scene segmentation,'' in \emph{Proceedings of the IEEE/CVF
  Conference on Computer Vision and Pattern Recognition}, 2019, pp. 3146--3154.

\end{thebibliography}
\end{document}